\documentclass[12pt,preprint]{aastex}

\usepackage{lscape}

\slugcomment{}

\shorttitle{Intermediate-Scale Anisotropy of Cosmic Rays Measured with the Telescope Array}
\shortauthors{The Telescope Array Collaboration}

\begin{document}


\title{Indications of Intermediate-Scale Anisotropy of Cosmic Rays with \\ Energy Greater Than 57~EeV in the Northern Sky Measured \\ with the Surface Detector of the Telescope Array Experiment}



\author{
R.U.~Abbasi$^{1}$,
M.~Abe$^{13}$,
T.Abu-Zayyad$^{1}$, 
M.~Allen$^{1}$, 
R.~Anderson$^{1}$, 
R.~Azuma$^{2}$, 
E.~Barcikowski$^{1}$, 
J.W.~Belz$^{1}$, 
D.R.~Bergman$^{1}$, 
S.A.~Blake$^{1}$
% ...
}
\affil{
$^{1}$ High Energy Astrophysics Institute and Department of Physics and Astronomy, University of Utah, Salt Lake City, Utah, USA \\
$^{2}$ Graduate School of Science and Engineering, Tokyo Institute of Technology, Meguro, Tokyo, Japan \\
% ...
$^{13}$ The Graduate School of Science and Engineering, Saitama University, Saitama, Saitama, Japan \\
% ...
}


\begin{abstract}
We have searched for intermediate-scale anisotropy 
in the arrival directions of ultrahigh-energy
cosmic rays with energies above 57~EeV in the northern sky
using data collected over a 5 year period by the 
surface detector of the Telescope Array experiment. 
We report on a cluster of events that we call the hotspot, 
found by oversampling using 20$\degr$-radius circles.
The hotspot has a Li-Ma statistical significance of 5.1$\sigma$, 
and is centered at ${\rm R.A.}=146\fdg7$, ${\rm Dec.}=43\fdg2$.  
The position of the hotspot is about 19$\degr$ off of the 
supergalactic plane. 
The probability of a cluster of events of 5.1$\sigma$ significance, 
appearing by chance in an isotropic cosmic-ray sky, is estimated to be 3.7$\times$10$^{-4}$
(3.4$\sigma$).
\end{abstract}





\section{Introduction}

The origin of ultrahigh-energy cosmic rays (UHECRs), 
particles with energies greater than 10$^{18}$~eV,
is one of the mysteries of astroparticle physics.
Greisen, Zatsepin, and Kuz'min (GZK)
predicted that UHECR protons with energies greater than 
$\sim$60~EeV ($6\times10^{19}$~eV) 
would be severely attenuated primarily due to pion photoproduction interactions 
with the cosmic microwave background (CMB) radiation \citep{Gre66,Zat66}. 
This GZK suppression becomes strong if these very high energy cosmic
rays are produced at and traveling moderate extragalactic distances.
The High Resolution Fly's Eye (HiRes) collaboration was first to
observe a suppression of cosmic rays above $\sim$60~EeV \cite{Abb08}, which is consistent with expectation from the GZK cutoff.
This suppression was independently confirmed by both the Pierre Auger Observatory (PAO) \citep{Abr08}
in the south and Telescope Array (TA) experiment \citep{Abu13a} in the north, which are the
largest aperture cosmic-ray detectors currently in operation. 

The distribution of UHECR sources 
should be limited within the local universe with distances smaller than 
100~Mpc for proton/iron and 20~Mpc for helium/carbon/nitrogen/oxygen
(distances within which $\sim$50\% of cosmic rays are estimated to survive) \citep{Kot11}.
To accelerate particles up to the ultrahigh-energy region,
particles must be confined to the accelerator site for more than a
million years by a magnetic field and/or a large-scale confinement volume
\citep{Hil84,Pti10}. This would thus limit the number of possible
accelerators in the universe to astrophysical
candidates such as galaxy clusters, supermassive black holes in active galactic nuclei (AGNs), 
jets and lobes of active galaxies, starburst galaxies, gamma-ray bursts, and magnetars. Galactic
objects are not likely to be the sources since
past observations indicate that the UHECRs do not concentrate in the galactic plane and have a relatively
isotropic distribution.  In addition, our galaxy cannot confine UHECRs above 10$^{19}$~eV within its volume by the Galactic magnetic field. 
Extragalactic astrophysical objects form
well-known large-scale structures (LSSs), most of which are 
spread along the ``supergalactic plane'' in the local universe. Nearby
AGNs are clustered and concentrated around LSS with a typical
clustering length of 5--15~Mpc, as observed by Swift BAT
\citep{Cap10}. 
Concentrations of nearby AGNs coincide
spatially with the LSS of matter in the local universe, including
galaxy clusters such as Centaurus and Virgo. 
The typical amplitude of such AGN concentrations is
estimated to be a few hundred percent of the averaged density within
a 20$\degr$-radius circle, which is of an angular scale comparable to
the clustering length of the AGNs within 85~Mpc \citep{Aje12}.

The main difficulty in identifying the origin of UHECRs is the loss
of directional information due to magnetic field induced bending.
In order to investigate the UHECR propagation from the extragalactic sources,
a number of numerical simulations have been developed
\citep{Yos03,Sig04,Taka06,Kas08,Koe09,Taka10,Kal11,Taka12}. In the
simulations, the UHECR trajectory between the assumed UHECR source
and the Earth is traced through intergalactic and galactic
magnetic fields (IGMF and GMF). The results depend strongly on
the assumed distribution and density of the UHECR sources and the
intervening magnetic fields. 
The deflection angle of a 60~EeV proton from a source at a distance of 50~Mpc is estimated to be a few degrees assuming models with an 
IGMF strength of 1~nG. 
Meanwhile, the estimated deflection by the 
GMF ranges from a few to about 10 degrees.
This, however, depends on the direction in the sky. 
If the highest-energy cosmic rays come from 
the local universe such as nearby galaxies, and if they are
protons, the maximum amplitude of the cosmic-ray anisotropy 
above $\sim$60~EeV is expected to be a few hundred percent of the 
average cosmic-ray flux. In this case, the amplitude of the cosmic-ray
anisotropy might be detectable by the UHECR detectors of
the TA and PAO.

In the highest-energy region, $E>57$~EeV, the PAO found correlations
of the cosmic-ray directions within a 3$\fdg$1-radius circle centered
at nearby AGNs (within 75~Mpc) in the southern sky \citep{Abr07}.
Updated measurements from
the PAO indicate a weakened correlation with nearby AGNs 
\citep{Abr10,Mac12}; the correlating fraction (the number of
correlated events divided by all events) decreased from the early
estimate of ($69^{+11}_{-13}$)\% to ($33\pm5$)\%, compared with
21\% expected for an isotropic distribution of cosmic rays. 
The chance probability of the original (69\%) correlation is 
$6\times10^{-3}$ assuming an isotropic sky. 
The Telescope Array
has also searched for UHECR anisotropies such as autocorrelations,
correlations with AGNs, and correlations with the LSS of the
universe using the first 40 months of scintillator surface detector (SD) data \citep{Abu12b,Abu13b}. 
Using 5 years of SD data, we updated results of the cosmic-ray anisotropy 
with $E>57$~EeV, which shows deviations from isotropy at the significance
of 2--3$\sigma$ \citep{Fuk13}. 
In this letter, we report on indications of intermediate-scale anisotropy of
cosmic rays with $E>57$~EeV in the northern hemisphere sky using the 5-year TA SD dataset.

\section{Experiment}

The Telescope Array is the largest cosmic-ray detector in the northern hemisphere.
It consists of a scintillator surface detector 
(SD) array \citep{Abu12a} and three fluorescence detector (FD) stations \citep{Tok12}. 
The observatory has been in full operation in Millard Country, Utah, USA
($39\fdg30$N, $112\fdg91$W; about 1,400~m above sea level) since
2008. The TA SD array consists of 507 plastic scintillation
detectors each 3~m$^2$ in area and located on a 1.2~km square grid.
The array has an area of $\sim$700~km$^{2}$. 
The TA SD array observes cosmic ray induced extensive air showers with $E>\sim$1~EeV, regardless of weather
conditions with a duty cycle near 100\% and a wide field of view (FoV). These capabilities ensure a very stable and large geometrical exposure over the northern sky survey in comparison with FD observations that
have a duty cycle of $\sim$10\%. 

\section{Dataset}

In this analysis, we used 
SD data recorded between 2008 May 11 and 2013 May 4.
The dataset contains approximately 1 million 
triggered events. 
For the reconstructed events, the energies determined by the 
SD array were renormalized by 1/1.27 to match the SD energy scale 
to that of the FD, which was determined calorimetrically \citep{Abu13a}.
Of these events, 72 met the
following conditions: (1) each event included at least four
SD counters; (2) the zenith angle of the event arrival direction was
less than 55$\degr$; and (3) the reconstructed energy
was greater than 57~EeV, which corresponds to the energy threshold
determined from the AGN correlation analysis results obtained by the
PAO \citep{Abr07}, and is adopted here to avoid introducing a free parameter
in the scanning phase space. 

The event selection criteria above are somewhat looser than those of our 
previous analyses of cosmic-ray anisotropy \citep{Fuk13}
to increase the observed cosmic-ray statistics. In our previous analyses, 
the largest signal counter is surrounded by 4 working counters that are its nearest 
neighbours to maintain the quality of the energy resolution and angular resolution. 
Only 52 events survived those tighter cuts.
When, the edge cut is abolished from the analysis (presented here)
to keep more cosmic-ray events, 20 events with $E>57$~EeV 
are recovered compared with the tighter cut analysis.
A full Monte Carlo (MC) simulation, which includes detailed detector responses \citep{Abu13a}, predicted a 13.2 event increase in the number of events. 
The chance probability of the data increment being 20 as compared to the MC prediction of 13.2 is estimated to be 5\%, which is within the range of statistical fluctuations.
The angular resolution of array boundary events
deteriorates to $1\fdg7$, compared to $1\fdg0$ for the well contained events.  
The energy resolution of array boundary
events also deteriorates to $\sim$20\%, where that of the inner array events is $\sim$15\%. 
These resolutions are still good enough to
search for intermediate-scale cosmic-ray anisotropy.  One final check is that when we calculate the cosmic ray
spectrum using the loose cuts analysis, the result is consistent with our published spectrum.

\section{Results}

\begin{figure}
\epsscale{1.00}
\plotone{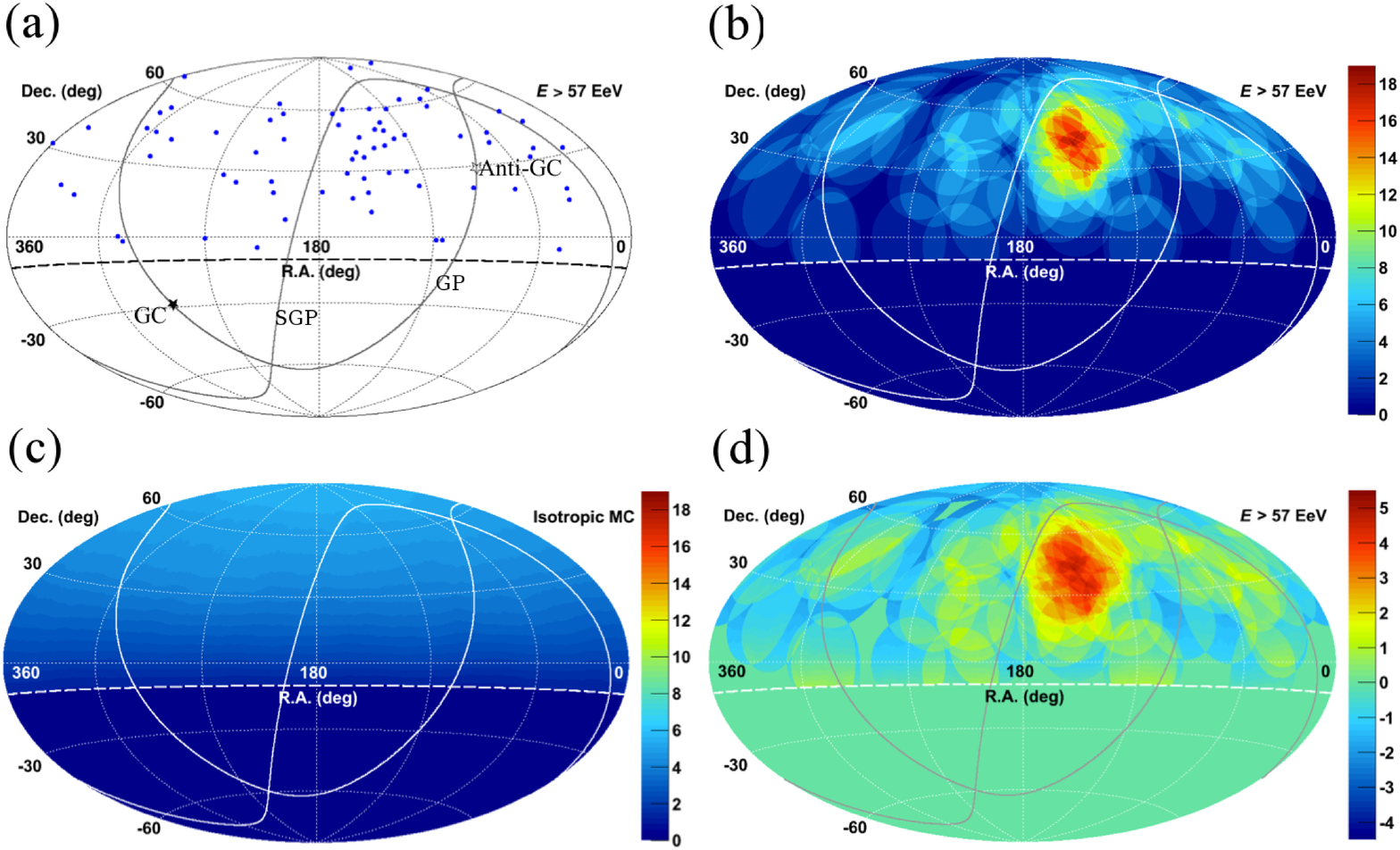}
\caption{Aitoff projection of the UHECR maps in equatorial
  coordinates.  The solid curves indicate the galactic plane (GP) and
  supergalactic plane (SGP). Our FoV is defined as the region above
  the dashed curve at ${\rm Dec.}=-10\degr$. (a) The points show the
  directions of the UHECRs $E>57$~EeV observed by the TA SD array,
  and the closed and open stars indicate the Galactic center (GC) and
  the anti-Galactic center (Anti-GC), respectively; (b) color contours
  show the number of observed cosmic ray events summed over a 20$\degr$-radius circle; (c)
  number of background events from the geometrical exposure summed over a
  20$\degr$-radius circle (the same color scale as (b) is used for
  comparison); (d) significance map calculated from (b) and (c) using
  Equation~\ref{Eq1}.} \label{fig1}
\end{figure}

Figure~\ref{fig1} (a) shows a sky map in equatorial coordinates of the
72 cosmic-ray events with energy $E>57$~EeV observed by the TA SD
array.  A cluster of events appears in this map centered near right
ascension $\sim$150$\degr$, and declination $\sim$40$\degr$, with a
diameter of $\sim$30$\degr$--40$\degr$.  In order to determine the
characteristics of the cluster, and estimate the significance of this
effect, we choose to apply elements of an analysis that was developed
by the AGASA collaboration to search for large-size anisotropy \citep{Hay99a,Hay99b}, 
namely to use oversampling with a 20$\degr$ radius.
Being mindful that scanning the parameter space of the analysis causes 
a large increase in chance corrections, we have not varied this radius.  
The TA and HiRes collaborations used this method previously \citep{Kaw13,Iva07}
to test the AGASA intermediate-scale anisotropy results with their data in the
10$^{18}$~eV range. 
The present letter reports on an extension of this method 
with application to the $E>57$~EeV energy region.

In our analysis, at each point in the sky map, cosmic ray events are summed over a
20$\degr$-radius circle as shown in Figure~\ref{fig1} (b).
The centers of tested directions are on a $0\fdg1
\times 0\fdg1$ grid from $0\degr$ to $360\degr$ in right ascension
(R.A.) and $-10\degr$ to $90\degr$ in declination (Dec.).  
We found that the maximum of $N_{\rm on}$, the number of observed events in a circle
of 20$\degr$ radius is $19$ within the TA FoV.
To estimate the number of background events under the signal 
in $N_{\rm on}$, we generated 
100,000 events assuming an isotropic flux.
We used a geometrical exposure $g(\theta)={\rm sin}\theta{\rm cos}\theta$
as a function of zenith angle ($\theta$) because the detection 
efficiency above 57~EeV is $\sim$100\%.
The zenith angle distribution deduced from the geometrical exposure is
consistent with that found in a full MC simulation.
The MC generated events are summed over each
20$\degr$-radius circle in the same manner as the data analysis, and
the number of events in each circle is defined as $N_{\rm
off}$. Figure~\ref{fig1} (c) shows the number of background events
$N_{\rm bg} = \eta N_{\rm off}$, where $\eta = 72/100,000$ is the
normalization factor.

We calculated the statistical significance of the excess
of events compared to the background events 
at each grid point of sky using the following equation
\citep{Li83}:
\begin{equation}\label{Eq1}
S_{\rm LM} = \sqrt{2}\left[N_{\rm on}~{\rm ln}\left(
\frac{(1+\eta)N_{\rm on} }{\eta (N_{\rm on} + N_{\rm off})}\right) +
N_{\rm off}~{\rm ln}\left(\frac{(1+\eta) N_{\rm off} }{N_{\rm on} +
N_{\rm off}} \right)\right]^{1/2}.
\end{equation}
Figure~\ref{fig1} (d) shows a significance map (in equatorial
coordinates) of the events above 57~EeV as observed by the TA SD
array. The maximum excess in our FoV appears as a
``hotspot'' centered at R.A.($\alpha$) $=146\fdg7$, Dec.($\delta$)
$=43\fdg2$ with a statistical significance of $S_{\rm MAX} =
5.1\sigma$ ($N_{\rm on} = 19, N_{\rm bg} = 4.49$).

The significance of the hotspot, quoted above at 5.1$\sigma$, does
not take random clustering into account, so one must make a
correction.  We did not carry out a blind analysis, but have been
watching the hotspot grow over several years as we collected further data and
added events to the sky plot. It is difficult to estimate the penalty due to
our having seen the cluster of events.  For example, in applying the oversampling
technique used by the AGASA experiment, we knew the oversampling radius roughly
matched the size of the hotspot cluster. 

However, by making a simple MC calculation one can estimate the probability of such a hotspot
appearing by chance anywhere in an isotropic sky.  One generates many isotropic MC event sets, each with the statistics 
of the experimental data, then performs a calculation of the Li-Ma significance exactly as was done on the data; i.e., using
oversampling with a radius of 20 degrees.  One can go further and approximate the effect of the eye's estimate of the radius of 
the cluster of events by repeating the calculation at other oversampling radii.  We did this, choosing five oversampling radii, 
15, 20, 25, 30, and 35 degrees.  We chose a 5 degree scan since by eye one cannot make an estimate more accurately than 
about $\pm$5 degrees. 

\begin{figure}
\epsscale{0.80}
\plotone{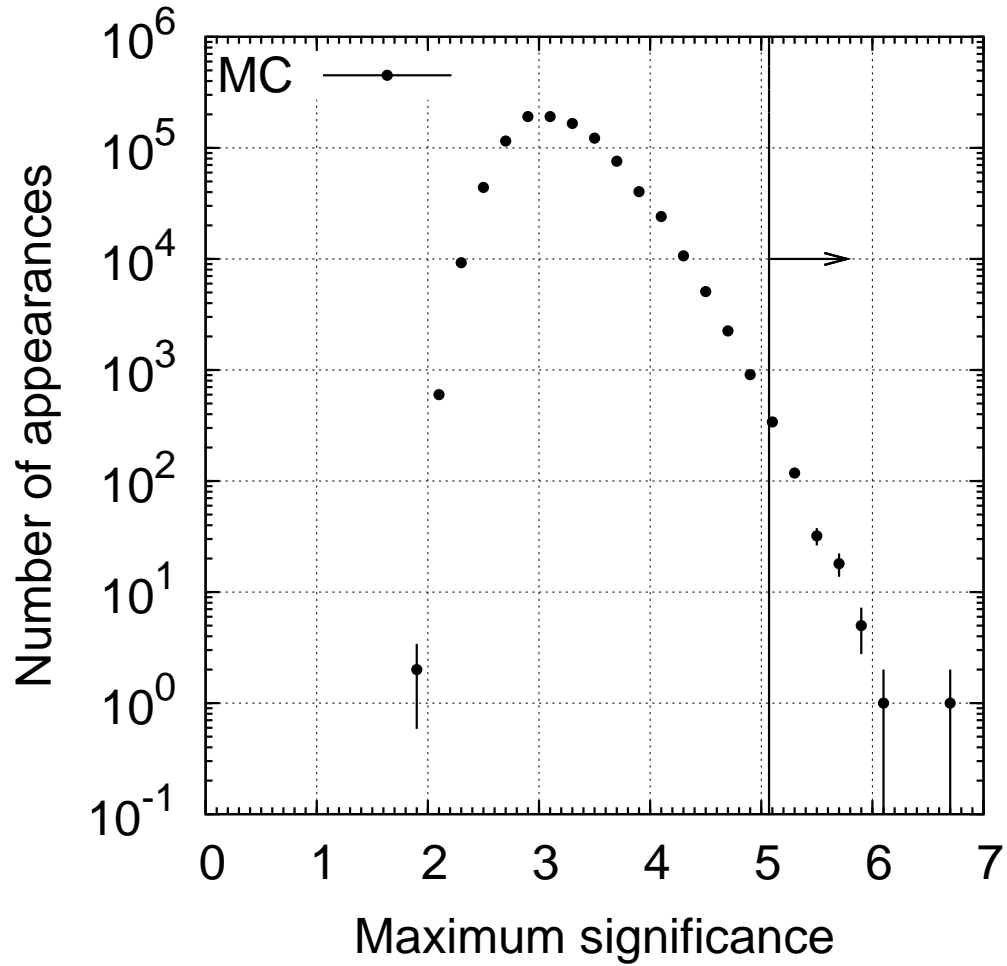}
\caption{Distribution of the maximum significance in our FoV as
  determined by a simple MC simulation assuming an isotropic flux. In the
  set of 1 million trials, each with 72 events, there were 365 instances of $S_{\rm MAX} > 5.1\sigma$.  This is indicated by the solid line and arrow in the plot. } \label{fig2}
\end{figure}

We generated 1 million MC data sets, each having 72 spatially random events within our FOV 
(i.e., we reproduced the statistics of the experimental data), assuming a uniform distribution over the TA SD exposure.
The maximum of the significances, $S_{\rm MAX}$, was calculated for each MC dataset 
in the same way as in the data, with the exception that the five oversampling radii were used, and the largest $S_{\rm MAX}$
was chosen. The distribution of the largest $S_{\rm MAX}$ of the 1 million
datasets is shown in Figure~\ref{fig2}.
We found that there were 365 instances of
$S_{\rm MAX}>5.1\sigma$. This yields a chance probability of the
observed hotspot in an isotropic cosmic-ray sky of
3.7$\times$10$^{-4}$, equivalent to a one-sided probability of 3.4$\sigma$.

\begin{figure}
\epsscale{0.80}
\plotone{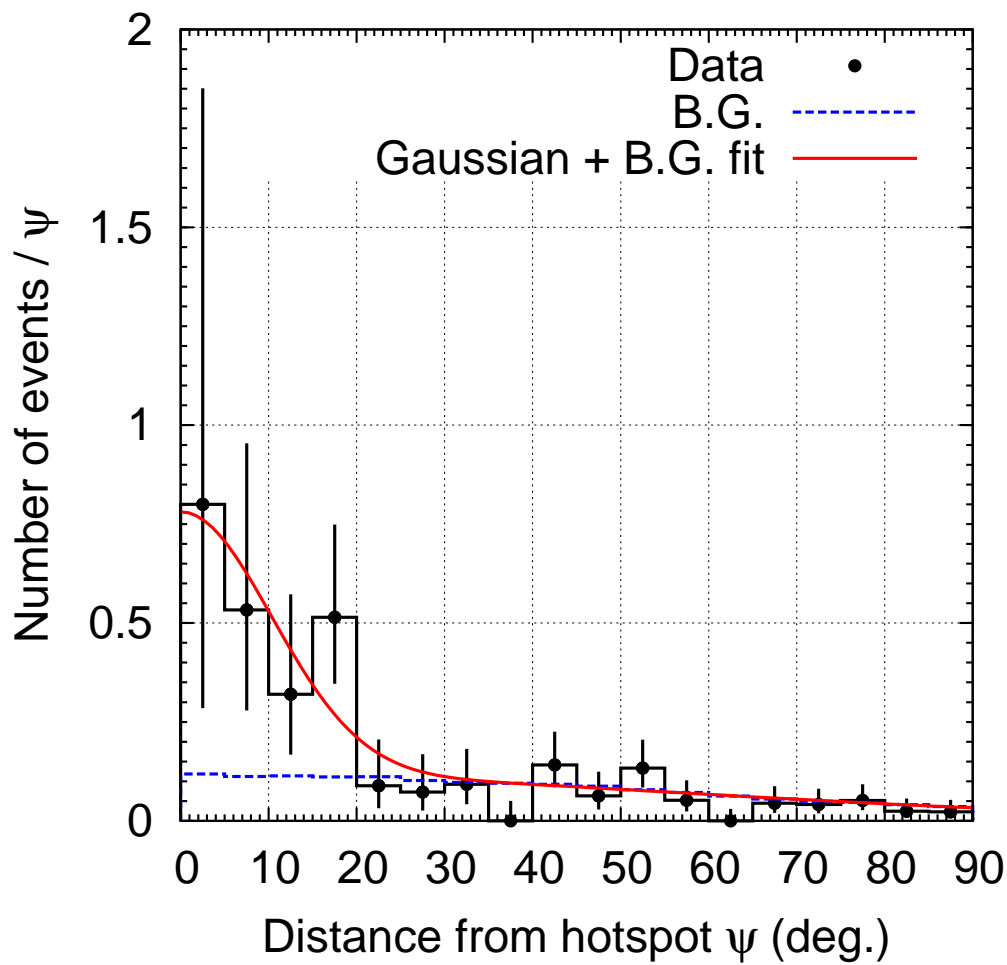}
\caption{Normalized number of events as a function of the opening
  angle ($\psi$) relative to the hotspot. The histogram shown in black
  with the points shows the observed events above 57~EeV by the TA
  SD. The dashed blue histogram shows the background events calculated
  by the MC simulation. The solid red curve is a fit using the binned
  maximum likelihood method and a Gaussian function with the
  background calculated with Equation~\ref{Eq2}. The signal spread and
  height are estimated to be $\sigma_{\rm s} = 10\fdg3 \pm 1\fdg9$ and
  $A_{\rm s} = 0.67 \pm 0.29$, respectively, assuming a Gaussian
  shape. } \label{fig3}
\end{figure}

To estimate the size of the hotspot, we present (see Figure~\ref{fig3}) the normalized number of events as
a function of the opening angle, $\psi$, relative to the center of the hotspot in the data.  Although with current
statistics we cannot determine the shape of the hotspot, to estimate its overall size we fit the hotspot excess using the 
binned maximum likelihood method, assuming a Gaussian signal plus a background estimated by the MC
simulation.  We used the following equation:
\begin{equation}\label{Eq2}
f(\psi; A_{\rm s}, \sigma_{\rm s}) = A_{\rm s}~{\rm
exp}\left(-\frac{\psi^{2}}{2\sigma_{\rm s}^{2}}\right) + (a_{0} +
a_{1}\psi^{2} + a_{2}\psi^{4}),
\end{equation}
where the first term is the Gaussian signal, and $A_{\rm s}$ and
$\sigma_{\rm s}$ denote fitting parameters of the signal height and spread,
respectively. The second term is the shape of the background
fitted by a polynomial function determined from the MC simulations
($a_{0} = 0.118$, $a_{1} = -1.7\times10^{-5}$, and $a_{2} =
8.5\times10^{-10}$). The spread of the hotspot was 
$\sigma_{\rm s} = 10\fdg3\pm1\fdg9$ ($A_{\rm s} = 0.67\pm0.29$). 
The uncertainty in the position of the hotspot is estimated to be
$\sigma_{\rm s} / \sqrt{N_{\rm on} - N_{\rm bg}} = 2\fdg7$.

\section{Discussion}

There are no known specific sources behind the hotspot.
The hotspot is located near the supergalactic plane, which
contains local galaxy clusters such as the Ursa Major cluster (20~Mpc from Earth), the Coma cluster (90~Mpc), and the Virgo cluster (20~Mpc). The angular distance between the hotspot center and the supergalactic plane in the vicinity of the Ursa Major cluster is $\sim$19$\degr$.  

Assuming the hotspot is real, two possible interpretations are:  it may be associated with the closest
galaxy groups and/or the galaxy filament connecting us with the Virgo cluster \citep{Dol04}; or if cosmic
rays are heavy nuclei they may originate close to the supergalactic plane, and be deflected by extragalactic magnetic fields and the galactic halo field \citep{Tin02,Taka12}.  To determine the origin of the hotspot, we will need greater UHECR
statistics in the northern sky. Better information about the mass composition of the
UHECRs, GMF, and IGMF would also be important.

\section{Summary}

Using cosmic ray events with energy $E>57$~EeV, collected over 5 years with the TA SD, we have
observed a cluster of events, which we call the hotspot, with a statistical significance of
5.1$\sigma$ ($N_{\rm on} = 19, N_{\rm bg} = 4.49$), 
centered at ${\rm R.A.}=146\fdg7$, ${\rm Dec.}=43\fdg2$.  
We calculated the probability of such a hotspot appearing 
by chance in an isotropic cosmic-ray sky to be 
3.7$\times$10$^{-4}$ (3.4$\sigma$).  

This indication of intermediate-scale anisotropy is limited by statistics collected by 
experiments in the northern hemisphere.  It provides a strong impetus for an improved effort 
to study the origin of UHECRs.  The TA$\times$4 project (extension of the TA SD by a factor
of 4) \citep{Sag13} is designed to provide the equivalent of 
20 TA-years of SD data by 2019, which would yield a 
$\sim$7 $\sigma$ observation if the ratio of hotspot to background events remains as is currently seen.  
TA$\times$4 and other related projects will enable us to make a precise UHECR anisotropy map 
with high statistics and help solve the mystery of the UHECR origin.

\acknowledgments The Telescope Array experiment is supported by the
Japan Society for the Promotion of Science through Grants-in-Aids for
Scientific Research on Specially Promoted Research (21000002)
``Extreme Phenomena in the Universe Explored by Highest Energy Cosmic
Rays'' and for Scientific Research (19104006), and the
Inter-University Research Program of the Institute for Cosmic Ray
Research; by the U.S. National Science Foundation awards PHY-0307098,
PHY-0601915, PHY-0649681, PHY-0703893, PHY-0758342, PHY-0848320,
PHY-1069280, and PHY-1069286; by the National Research Foundation of
Korea (2007-0093860, R32-10130, 2012R1A1A2008381, 2013004883); by the
Russian Academy of Sciences, RFBR grants 11-02-01528a and 13-02-01311a
(INR), IISN project No. 4.4509.10 and Belgian Science Policy under
IUAP VII/37 (ULB). The foundations of Dr. Ezekiel R. and Edna Wattis
Dumke, Willard L. Eccles and the George S. and Dolores Dore Eccles all
helped with generous donations. The State of Utah supported the
project through its Economic Development Board, and the University of
Utah through the Office of the Vice President for Research. The
experimental site became available through the cooperation of the Utah
School and Institutional Trust Lands Administration (SITLA),
U.S. Bureau of Land Management, and the U.S. Air Force. We also wish
to thank the people and the officials of Millard County, Utah for
their steadfast and warm support. We gratefully acknowledge the
contributions from the technical staffs of our home institutions. An
allocation of computer time from the Center for High Performance
Computing at the University of Utah is gratefully acknowledged.






\appendix

\section{List of Events with $E>57$~EeV}
In this Appendix we present the list of events with energy $E>57$~EeV
and zenith angle $\theta < 55\degr$ that have been recorded by the surface
detector of the Telescope Array from May 11, 2008 to May 4,
2013. During this period, 72 such events were observed. Table~\ref{tbl-1} shows
the arrival date and time of these events, the zenith angle, energy in
units of EeV, and equatorial coordinates $\alpha$ (R.A.) and $\delta$ (Dec.) in degrees.
Noted that the air shower reconstruction used here as described in Kawata et al. (2013)
was slightly different from that of previous anisotropy work \citep{Abu12b}. 
The opening angles between these arrival directions and previous ones
are almost within 1$\degr$. This difference hardly affects the results of the large-scale anisotropy.

\clearpage

\clearpage

\clearpage

\begin{deluxetable}{lrrrr}
\tabletypesize{\scriptsize}
\tablecaption{List of Telescope Array events with $E>$57~EeV and zenith angle $\theta<55\degr$ recorded from 2008 May 11 to 2013 May 4 \label{tbl-1}}
\tablewidth{0pt}
\tablehead{
\colhead{Date and Time (UTC)} & \colhead{$\theta$ (deg)} & \colhead{$E$ (EeV)} & \colhead{$\alpha$ (deg)} & \colhead{$\delta$ (deg)} }
\startdata
2008 Jun 10 	17:05:37 &	 46.91 &	 88.8 &	  93.50 &   20.82 \\
2008 Jun 25 	19:45:52 &	 31.98 &	 82.6 &	  68.86 &   19.20 \\
2008 Jun 29 	08:22:45 &	 41.20 &	101.4 &	 285.74 &   $-$1.69 \\
2008 Jul 15 	05:26:31 &	 34.26 &	 57.3 &	 308.45 &   53.91 \\
2008 Jul 20 	04:35:32 &	 25.61 &	120.3 &	 285.46 &   33.62 \\
2008 Aug 01 	23:01:33 &	 39.43 &	139.0 &	 152.27 &   11.10 \\
2008 Aug 09 	06:16:16 &	 14.54 &	 76.9 &	 280.28 &   41.34 \\
2008 Aug 10 	12:45:04 &	 38.04 &	122.2 &	 347.73 &   39.46 \\
2008 Sep 24 	20:09:22 &	 23.16 &	 68.8 &	 178.03 &   20.29 \\
2008 Oct 08 	18:20:16 &	 24.69 &	 69.1 &	 154.49 &   26.50 \\
2008 Oct 30 	17:30:35 &	 27.45 &	 79.3 &	 152.44 &   45.84 \\
2008 Nov 08 	14:30:41 &	 15.16 &	 59.2 &	 139.66 &   28.72 \\
2008 Dec 30 	10:49:32 &	  4.11 &	 59.7 &	 152.30 &   36.45 \\
2009 Jan 22 	22:54:22 &	 31.27 &	 57.9 &	 311.15 &   51.06 \\
2009 Feb 12 	01:00:17 &	 41.63 &	 64.2 &	  22.49 &   80.07 \\
2009 Mar 28 	04:36:08 &	 34.48 &	 80.7 &	  99.16 &   62.77 \\
2009 Mar 29 	03:43:34 &	 20.84 &	 75.0 &	 119.62 &   59.19 \\
2009 Apr 28 	19:22:32 &	 20.68 &	 64.5 &	  61.60 &   42.91 \\
2009 May 19 	02:19:52 &	 42.51 &	 64.2 &	 206.74 &   24.93 \\
2009 Jun 14 	01:33:00 &	 52.23 &	 62.5 &	 235.00 &   27.63 \\
2009 Jul 13 	04:38:52 &	 40.41 &	154.3 &	 239.85 &   $-$0.41 \\
2009 Aug 14 	10:26:17 &	 45.02 &	 59.5 &	 305.06 &   44.36 \\
2009 Aug 15 	08:54:55 &	 23.27 &	 65.2 &	 331.65 &   18.85 \\
2009 Sep 19 	08:45:52 &	 34.07 &	 61.7 &	  56.47 &   64.36 \\
2009 Oct 21 	10:49:42 &	  4.38 &	 66.5 &	  82.19 &   43.12 \\
2009 Nov 26 	12:33:08 &	 16.61 &	 64.0 &	 120.03 &   46.05 \\
2010 Jan 08 	07:17:31 &	 19.04 &	 57.6 &	 128.77 &   44.51 \\
2010 Jan 21 	03:53:51 &	 23.55 &	 61.2 &	  78.79 &   61.43 \\
2010 Feb 22 	07:10:34 &	 11.58 &	 63.7 &	 139.13 &   49.62 \\
2010 Feb 26 	00:11:32 &	 15.83 &	 65.2 &	  25.28 &   43.99 \\
2010 Jul 11 	03:01:45 &	 44.90 &	 58.0 &	 212.37 &   $-$4.78 \\
2010 Jul 14 	17:06:43 &	 51.15 &	 92.2 &	 144.59 &   40.66 \\
2010 Aug 05 	19:44:07 &	 45.54 &	 67.1 &	 115.13 &   $-$1.45 \\
2010 Aug 29 	21:20:45 &	 36.12 &	 68.9 &	 137.13 &   41.50 \\
2010 Aug 30 	20:50:45 &	 19.60 &	 93.5 &	 204.00 &   45.18 \\
2010 Sep 16 	20:26:32 &	 49.72 &	 60.5 &	 129.28 &   29.13 \\
2010 Sep 19 	07:05:00 &	 23.66 &	 66.3 &	  19.29 &   32.26 \\
2010 Sep 21 	20:37:06 &	 20.56 &	162.2 &	 205.08 &   20.05 \\
2011 Jan 05 	00:56:23 &	  8.93 &	 67.4 &	 359.91 &   31.47 \\
2011 Jan 08 	18:40:41 &	 15.88 &	124.8 &	 295.61 &   43.53 \\
2011 Feb 28 	16:16:26 &	 38.94 &	135.5 &	 288.30 &    0.34 \\
2011 Apr 17 	20:20:29 &	 34.09 &	 74.7 &	  82.50 &   57.70 \\
2011 Jul 13 	19:12:34 &	 42.73 &	 65.4 &	  87.59 &   81.53 \\
2011 Jul 18 	22:16:57 &	 54.23 &	 73.9 &	 118.41 &   $-$1.37 \\
2011 Jul 22 	22:15:41 &	 10.64 &	 62.3 &	 163.67 &   28.92 \\
2011 Jul 24 	23:17:22 &	 35.99 &	 61.2 &	 197.78 &    7.74 \\
2011 Jul 28 	15:21:08 &	 19.18 &	 89.3 &	  39.98 &   34.17 \\
2011 Aug 21 	04:52:22 &	 48.34 &	 69.2 &	 218.81 &   54.11 \\
2011 Aug 25 	22:05:12 &	 23.98 &	 83.3 &	 168.48 &   57.92 \\
2011 Aug 28 	21:14:19 &	 31.69 &	 63.3 &	 153.21 &   19.83 \\
2011 Sep 10 	21:16:00 &	 44.24 &	 78.8 &	 133.62 &   48.55 \\
2011 Oct 03 	17:23:40 &	 22.03 &	 72.6 &	 161.74 &   17.39 \\
2011 Oct 21 	06:33:06 &	 15.71 &	 78.7 &	  31.32 &   49.49 \\
2011 Oct 22 	21:23:09 &	 12.68 &	 57.6 &	 253.12 &   46.43 \\
2011 Nov 09 	16:51:22 &	 24.30 &	 72.9 &	 156.84 &   38.82 \\
2011 Nov 17 	14:50:55 &	 26.16 &	 81.6 &	 132.97 &   52.63 \\
2011 Nov 19 	22:39:41 &	 38.07 &	 57.4 &	 319.95 &   15.87 \\
2012 Jan 05 	16:31:21 &	 18.36 &	 91.8 &	 226.68 &   24.50 \\
2012 Mar 13 	02:54:54 &	 25.25 &	 60.3 &	 123.94 &   22.52 \\
2012 Apr 11 	06:20:45 &	 29.10 &	101.0 &	 219.66 &   38.46 \\
2012 May 04 	00:18:13 &	 24.42 &	 76.9 &	 134.84 &   59.83 \\
2012 Aug 19 	00:11:20 &	 18.78 &	 75.6 &	 210.35 &   57.55 \\
2012 Sep 07 	22:30:26 &	 38.79 &	 57.8 &	 158.60 &   60.26 \\
2012 Sep 18 	09:00:34 &	 28.61 &	 59.0 &	 355.95 &   64.19 \\
2012 Sep 27 	13:51:05 &	 45.68 &	 57.4 &	 159.75 &   35.56 \\
2012 Nov 01 	09:06:32 &	 46.60 &	 60.5 &	  47.71 &   $-$4.66 \\
2013 Jan 03 	09:56:20 &	 54.73 &	 68.2 &	  66.42 &   39.00 \\
2013 Mar 14 	23:56:46 &	 28.99 &	 98.5 &	  36.26 &   17.87 \\
2013 Mar 21 	22:45:18 &	 26.91 &	106.8 &	  37.59 &   13.89 \\
2013 Apr 08 	03:55:10 &	 49.65 &	 66.8 &	 218.49 &   62.54 \\
2013 Apr 22 	01:13:31 &	 36.20 &	 62.5 &	 165.28 &   52.35 \\
2013 Apr 28 	17:05:20 &	 38.73 &	 68.5 &	  47.08 &   31.32 \\
\enddata
\tablecomments{Table \ref{tbl-1} is published in its entirety in the 
electronic edition of the {\it Astrophysical Journal}.  A portion is 
shown here for guidance regarding its form and content.}
\end{deluxetable}







\end{document}